\documentclass[usenatbib]{mn2e}

\newcommand{\edit}{}
\newcommand{\redit}{}

\usepackage{graphicx,hyperref,amsmath,bm,xspace}
\hypersetup{colorlinks,linkcolor=blue, citecolor=blue}
\usepackage[compress]{cleveref}
\crefname{equation}{equation}{equations}
\crefname{figure}{Figure}{Figures}
\crefname{section}{section}{sections}
\crefname{subsection}{section}{sections}
\crefname{subsubsection}{section}{sections}
\crefname{table}{Table}{Tables}
\usepackage{etoolbox}

\newcommand{\tbhns}{$\tau_{\rm BH+NS}$\xspace}
\newcommand{\pbhns}{$p_{\rm BH+NS}$\xspace}
\makeatletter

\patchcmd{\NAT@citex}
  {\@citea\NAT@hyper@{%
     \NAT@nmfmt{\NAT@nm}%
     \hyper@natlinkbreak{\NAT@aysep\NAT@spacechar}{\@citeb\@extra@b@citeb}%
     \NAT@date}}
  {\@citea\NAT@nmfmt{\NAT@nm}%
   \NAT@aysep\NAT@spacechar\NAT@hyper@{\NAT@date}}{}{}

\patchcmd{\NAT@citex}
  {\@citea\NAT@hyper@{%
     \NAT@nmfmt{\NAT@nm}%
     \hyper@natlinkbreak{\NAT@spacechar\NAT@@open\if*#1*\else#1\NAT@spacechar\fi}%
       {\@citeb\@extra@b@citeb}%
     \NAT@date}}
  {\@citea\NAT@nmfmt{\NAT@nm}%
   \NAT@spacechar\NAT@@open\if*#1*\else#1\NAT@spacechar\fi\NAT@hyper@{\NAT@date}}
  {}{}

\makeatother

\newcommand{\msun}{\ensuremath{\;{\rm M_{\sun}}}\xspace}

\begin{document}

\title[Dynamically formed BH+MSP binaries]{Dynamically formed black hole+millisecond pulsar binaries in globular clusters}
\author[Clausen, Sigurdsson, \& Chernoff]{Drew Clausen,$^{1,2}$\thanks{Email: dclausen@tapir.caltech.edu} Steinn Sigurdsson,$^2$ David F. Chernoff $^3$\\
$^{1}$TAPIR, California Institute of Technology, Mailcode 350-17, Pasadena, CA 91125, USA\\
$^{2}$Department of Astronomy \& Astrophysics, The Pennsylvania State University, 525 Davey Lab, University Park, PA 16802, USA\\
$^{3}$Department of Astronomy, Cornell University, 610 Space Sciences Building, Cornell University, Ithaca, NY 14853, USA}

\date{Accepted 2014 April 29.  Received 2014 April 25; in original form 2014 March 13}
\maketitle

\begin{abstract}
The discovery of a binary comprising a black hole (BH) and a millisecond pulsar (MSP) would yield insights into stellar evolution and facilitate  {\edit exquisitely sensitive} tests  of general relativity.  Globular clusters (GCs) are known to harbor large MSP populations and recent studies suggest that GCs may also retain a substantial population of stellar mass BHs.  We modeled the formation of BH+MSP binaries in GCs through exchange interactions between binary and single stars.  We found that in dense, massive clusters most of the dynamically formed BH+MSP binaries will have orbital periods of 2 to 10 days, regardless of the mass of the BH, the number of BHs retained by the cluster, and the nature of the GC's binary population.  The size of the BH+MSP population is sensitive to several uncertain parameters, including the BH mass function, the BH retention fraction, and the binary fraction in GCs.  Based on our models, we estimate that there are $0.6\pm0.2$ dynamically formed BH+MSP binaries in the Milky Way GC system, and place an upper limit on the size of this population of $\sim 10$.  Interestingly, we find that BH+MSP binaries will be rare even if GCs retain large BH populations.        
\end{abstract}


\section{Introduction}
\label{sec:intro}
Radio pulsars in binary systems provide constraints on the processes that drive binary stellar evolution and unparalleled tests of general relativity in the strong-field regime.  In most cases, the pulsars in these binaries are ``recycled."  That is, the neutron star (NS) has been spun-up by accreting mass and angular momentum from its companion \citep{Alpar:1982}.  Compared to ``normal'' pulsars, recycled pulsars exhibit greater stability and have much shorter spin periods ($P_{\rm S} \la 100$ ms), both of which facilitate high precision measurements of the pulse arrival times  \citep{Lorimer:2008}.  {\edit Recycled pulsars with $P_{\rm S} \la 20$ ms are referred to as millisecond pulsars (MSPs).}  The outcomes of binary evolution can be probed by using the recycled pulsar in such a system as  a stable clock to precisely determine the binary's Keplerian orbital parameters and the properties of its component stars.  If the recycled pulsar's companion is another neutron star, it is possible to measure post-Keplerian orbital parameters in a model-independent fashion and then compare these measurements with the predictions of various theories of gravity \citep{Stairs:2004}.  The post-Keplerian parameters measured in the double pulsar binary PSR J0737-3039 offer the best test of gravity in the strong field limit, to date, and are in excellent agreement with the predictions of general relativity \citep{Kramer:2006}.  {\edit New and better tests} of general relativity may be possible by applying these techniques to a binary comprising a black hole (BH) and a recycled pulsar, however such a system has yet to be discovered. 

It is possible to produce a BH+recycled pulsar binary through standard evolutionary processes in an isolated, high-mass binary.  The scenario requires that the primary (the initially more massive member of the binary, which evolves faster than its companion) produces a NS at the end of its lifetime and that the secondary produces a BH.  This can occur if the primary  transfers enough material to its companion to drive the companion's mass above the threshold for BH production.  The NS created by the primary could then be recycled by accreting material from the companion before it evolves to become a BH.  Under the assumption that these recycled pulsars would have lifetimes longer than $10^{10}$ yr,  \citet{Narayan:1991} placed an empirical upper limit on the formation rate of such BH+recycled pulsar binaries of $10^{-6}~{\rm yr^{-1}}$ within the Galaxy.  Population synthesis models by \citet{Lipunov:1994} found a comparable formation rate, while similar studies by \citet{Sipior:2002}, \citet{Voss:2003}, \citet{Sipior:2004}, and \citet{Pfahl:2005} favored lower BH+recycled pulsar binary formation rates of $\sim 10^{-7}~{\rm yr^{-1}}$.  Additionally, \citet{Pfahl:2005} argued that the NSs in these systems would only be mildly recycled.  Due to the rapid evolution of its massive companion, the NS, accreting at the Eddington limit, would only have time to accrete $10^{-3}-10^{-4}\msun$ of material before the secondary collapsed into a BH.  These mildly recycled pulsars would only have lifetimes of $10^{8}$ yr.  Even if the pulsars in these systems were completely recycled {\edit into long-lived MSPs}, the population synthesis calculations suggest that most of these systems will undergo a gravitational wave driven merger within $\sim 10^{8}$ yr. {\edit With formation rates of $10^{-7}$  yr$^{-1}$ and lifetimes of $\la 10^8$ yr the number of BH+recycled pulsar binaries expected to exist in the Milky Way is only $\sim 10$.}  

In a globular cluster, a BH+recycled pulsar binary {\edit need not form directly from} a primordial binary.  The high stellar density in GCs leads to dynamical encounters between cluster members, which opens a wide array of evolutionary pathways that are inaccessible to isolated binaries.  For example, a single NS in a GC can gain a companion by exchanging into a primordial binary during a three-body interaction. Subsequent evolution of these newly created binaries can result in the NS being spun-up into a recycled, MSP (\citealt{Hills:1976,Sigurdsson:1995,Ivanova:2008}).  These encounters are evidenced by the enhanced formation rates of low mass X-ray binaries  (LMXBs) and their progeny, MSPs, observed in GCs \citep[e.g][]{Katz:1975, Verbunt:1987, Pooley:2003, Camilo:2005}.   Any of the BHs present in the cluster could acquire a MSP companion through similar interactions \citep{Sigurdsson:2003}.  However, uncertainties in the size and nature of the BH population present in GCs complicate investigations of this formation channel.  

\citet{Kulkarni:1993} and \citet{Sigurdsson:1993} argued that the stellar mass BHs formed in a GC would rapidly sink to the center of the cluster and eject one another in a phase of intense self-interaction. The frenzy of ejections results in a substantial {\edit depletion} of the cluster's stellar mass BH population during the first Gyr of evolution.  The fact that a firm BH candidate had not been identified in a GC during decades of observational study was inline with this theoretical picture.  Given the meager BH populations implied by these investigations, the dynamical formation of BH+MSP binaries in GCs has received little attention.  After all, this channel closes if there is {\edit no} population of BHs present in GCs.   Nevertheless, the production of BH+MSP binaries through multibody interactions has been considered in dense stellar environments, analogous to GCs, that are likely to harbor BHs.  \citet{Faucher-Giguere:2011} showed that a few dynamically formed BH+MSP binaries should be present in the Galactic Center, where a cluster of $\sim 10^{4}$ stellar mass BH is expected to exist.  This result indicates that BH+MSP binaries {\edit might be produced} in GCs if the clusters retained some of their stellar mass BHs.   

Recent observational efforts have shown that there are BHs present in some GCs, prompting a renewed interest in the nature of GC BH populations.  A number of promising BH candidates have been discovered in X-ray observations of extragalactic GCs \citep{Maccarone:2007, Maccarone:2011, Irwin:2010, Shih:2010, Brassington:2010, Brassington:2012, Roberts:2012, Barnard:2012}.  Furthermore, three BH candidates have been identified in deep radio observations of Milky Way GCs; two candidates reside in M22 and one candidate is in M62 \citep{Strader:2012,Chomiuk:2013}.  There is also a growing body of theoretical work suggesting that it may be possible for GCs to retain a substantial fraction of their stellar mass BH populations, under certain circumstances {\redit \citep{Mackey:2008,Sippel:2013,Morscher:2013,Breen:2013,Breen:2013a,Heggie:2014}}.  Motivated by these new results, we set out to explore how efficiently three-body exchanges produce BH+MSP binaries in GCs.    
   
It has also been suggested that GCs may harbor intermediate mass BHs (IMBHs; $M \sim 10^{2}-10^{4} \msun$).  Previous studies have considered the consequences of interactions between MSPs and these IMBHs. The encounters could result in a MSP being significantly displaced from the GC core \citep{Colpi:2003}, produce a IMBH+MSP binary \citep{Devecchi:2007}, or populate the Milky Way halo with several high velocity MSPs \citep{Sesana:2012}.  We will not include IMBHs in the models presented here, and instead focus on stellar mass BHs.

This paper is organized as follows.  We describe the features of our models and the motivate the range of initial conditions that our simulations explore in \cref{sec:method}.  In \cref{sec:orbparms} we discuss the orbital parameters of the BH+MSP binaries produced in our models.  We discuss the size of the BH+MSP binary population and the possibility of detecting such a binary in \cref{sec:detection}.  Finally, in \cref{sec:discussion}, we summarize and discuss our findings.             
 
\section{Method}
\label{sec:method}
Several characteristics of a GC can influence the nature of the BH+MSP binaries produced within it.  The observed diversity in GC structure combined with uncertainties in the clusters' binary and BH populations produce a vast range of plausible model inputs.  Exploring this parameter space efficiently necessitates the use of fast, approximate methods{\edit, which themselves require a substantial amount of computation time.  The study presented here} will motivate and inform observations necessary for constraining the parameter space, and identify interesting regions in this space for followup with detailed $N$-body or Monte Carlo calculations.    

We simulated the dynamical formation of BH+MSP binaries by evolving a variety of BH-binaries in fixed background GC models using the method described in \citet{Sigurdsson:1995} and \citet[][hereafter CSC13]{Clausen:2013}.  {\edit Here we briefly list the key physical effects included in our models:
\begin{itemize}
\item{Fokker--Planck advection and diffusion of single BHs or binaries containing at least one BH in static, multimass background clusters}
\item{direct integration of binary--single encounters to track changes in the orbital parameters of the binary, exchanges of members of the binary, destruction of the binary, and physical collisions {\redit between stars and/or BHs}}
\item{changes in the semimajor axis and eccentricity of the binary on account of gravitational wave emission}
\item{changes in the masses of binary components following mergers and recoil of BHs after gravitational wave driven mergers.} 
\end{itemize}   
We will describe our implementation of these effects in more detail below, but for a complete description we refer the reader to \citet{Sigurdsson:1995} and CSC13. }

{\redit We focus on multimass King models whose total stellar content derives from a simple, evolved mass function.  We start once the static, equilibrium approximation implicit in the King model treatment is reasonably accurate. The calculation begins only after stars of several solar masses have evolved off the main sequence because prior evolutionary phases involved significant mass loss. Collisional truncation of the wide binaries has already occurred. The cluster's BHs and NSs of interest in this paper were born much earlier.  Many important and complicated processes have run to completion, e.g. dynamical friction segregated heavy objects and promoted formation of a self-gravitating core. Interactions involving binaries have ejected all but a few of the heaviest remnants. We do not model any of these processes. We assume a priori (and without the guidance of direct observations) (1) the single versus binary content and (2) the number and mass of resident black holes. We focus on evolutionary stages when cluster changes occur on the characteristic half-mass relaxation time-scale. We specify structural properties that match observed GCs, apparently in the late time, stable phase seen in fully dynamical models of GC evolution \citep[e.g.][]{Hurley:2007,Chatterjee:2010}. These assumptions capture the outcomes of all the early, complicated phases of evolution and provide the initial conditions for our calculations.}  We generated four background cluster families to study the impact of GC structure on the formation rate and properties of the BH+MSP binaries.  The names, central number densities $n_{\rm c}$, mean central velocity dispersions $\bar{v}_{\rm m}$, concentrations $c$, and total masses $M_{\rm GC}$ of the these cluster families are listed in \cref{tab:gcparms}.  

Each cluster's stellar population was assumed to have an initial mass function of the form $\xi(m) \propto m^{-\alpha}$.  {\redit We used a broken power law similar to the one given in \citet{Kroupa:2001} with $\alpha$ values of $1.3$ and $2.35$ for stars with mass $m < 0.35~{\rm M_{\odot}}$ and $m > 0.35~{\rm M_{\odot}}$, respectively.}  We chose a main-sequence turn-off mass of $0.85~ {\rm M_{\odot}}$ and assumed that stars above the turn-off mass had evolved completely to white dwarfs (WDs), NSs, or BHs.  The evolved stellar population was then binned into 10 mass groups.  {\redit Main sequence stars and WDs with masses in the range 0.08 -- 1.2\msun were grouped into eight bins with widths of $\sim 0.15\msun$.  A ninth mass bin for stars with mass $1.2\msun<m<1.5\msun$ contained the NSs.  The final, high mass bin held the BHs.} WD masses were computed using the semi-empirical initial--final mass relation derived by \citet{Catalan:2008}.  We used a NS mass of $1.4~ {\rm M_{\odot}}$ and considered NS retention fractions, $f_{\rm ret}$, of 5\%, 10\%, and 20\% \citep{Sigurdsson:1995,Pfahl:2002,Ivanova:2008}.  

The BH populations in GCs are poorly constrained, so we considered a range of BH masses, $M_{\rm BH}$, and population sizes.  By analogy to the BHs found in the Galaxy, we used $M_{\rm BH} = 7 {\rm~M_{\odot}}$ \citep{Ozel:2010,Farr:2011}.   A second value of  $M_{\rm BH} = 15~{\rm M_{\odot}}$ was motivated by \citet{Strader:2012}, who used mass segregation arguments to estimate the masses of the BH candidates in M22 to be $\sim 15~{\rm M_{\odot}}$.  We also used $M_{\rm BH} = 35~{\rm M_{\odot}}$ in some of our simulations. Observations of the stellar mass BH candidates in extragalactic GC indicate that these BHs may have masses $\ga 30~{\rm M_{\odot}}$ \citep{Maccarone:2007,Barnard:2012,Irwin:2010,Clausen:2012}.  We controlled the number of BHs present in each simulation by truncating the high-mass end of the initial mass function.  In a subset of our simulations, we assumed that nearly every BH formed in the cluster was ejected, leaving a lone BH\footnotemark[1]. In other simulations, we allowed the clusters to retain several BHs, with the number of BHs ($N_{\rm BH}$) in the range $7-191$.  {\redit In such simulations, the additional BHs were included as part of the background cluster and comprised the high mass group of our multimass King model.  As such, all of these BHs were assumed to be single and held in mass-segregated equilibrium with the rest of the cluster.} Including these BH populations in our static background clusters had a substantial impact on the mass and structure of these clusters.  To facilitate comparisons between models with different values of $N_{\rm BH}$, we adjusted the free parameters in the cluster models, slightly, {\edit to ensure that $n_{\rm c}$, $\bar{v}_{\rm m}$, $c$, and $M_{\rm GC}$ changed by $\la 20\%$} when $N_{\rm BH}$ was varied.  The ranges of these parameters are given in \cref{tab:gcparms}.  
\footnotetext[1]{In CSC13 we discussed the BH+NS merger rates predicted by these simulations.}

{\edit We generated 2000 BH-binaries and followed their evolution, one-by-one,  in each of the background clusters.  We refer to this group of calculations as one simulation.}  The initial configuration of the BH-binary used in each run was determined using the following distributions.  The BH's companion was randomly selected from the cluster's evolved mass distribution.  We selected the initial eccentricity, $e$, from a thermal distribution, $f(e) \propto 2e$.  The distribution of initial semimajor axes was assumed to be flat in $\log a$ between $10^{-3}$ au and $a_{\rm max}$.  {\redit The initial BH-binaries in our simulations are the products of complicated stellar and dynamical evolution, the latter of which occurs at different rates in different clusters.  We set $a_{\rm max}$ to 100, 33, 15, and 10 au for simulations in cluster A, B, C and D, respectively, in an attempt to capture the broad range of the possible outcomes of this evolution {\edit(CSC13)}.}  Finally, the initial position and velocity of each BH-binary was selected from the radial density and velocity distributions of the third most massive mass group {\edit(CSC13)}.

\begin{table}
\centering
 \caption{Background GC Model Parameter Ranges\label{tab:gcparms}}
  \begin{tabular}{@{}ccccc@{}}
	\hline
	Family&$n_{\rm c}$& $\bar{v}_{\rm m}$&$M_{\rm GC}$ & $c$\\
	Name&$(10^{5}~{\rm pc^{-3}})$&$({\rm km\;s^{-1}})$& ($10^{5}~\rm{M_{\odot}}$)& \\
	\hline
	A&$0.1$&6&$1.0-1.2$&$1.20-1.35$\\
	B&$1.0-1.2$&$8.1-10$&$5.2-6.2$&$1.71-1.79$\\
	C&$5.0-5.2$&$10-11$&$7.2-8.5$&$1.93-2.02$\\
	D&$10$&11-13&$11-14$&$2.06-2.15$\\
	\hline
 \end{tabular}
 \end{table}

We sought to determine how often the BHs in these binaries acquired MSP companions.  {\edit Each binary was evolved} in the cluster potential with dynamical friction and Fokker-Planck diffusion calculated explicitly.  Each run continued until either the BH was ejected from the cluster or the run had covered a maximum time of $t_{\rm max} = 10^{10}$ yr. The probability that the binary would undergo a close encounter with a background star was calculated continuously (see \citealt{Sigurdsson:1995} for a detailed description).  {\redit In simulations of clusters that contained more than one BH, this included the possibility that the binary would interact with a background BH.  The probability of such an encounter was effectively 0 in the clusters that only contained one BH.}  If an encounter was deemed to have occurred, we integrated the three-body interaction between the binary and the background star.  These encounters can alter the orbital parameters and the components of the binary, due to a {\redit collision/merger} or an exchange.  The interaction may also result in the disruption of the binary.  If the binary survived the encounter, {\edit we continued to follow its evolution within the cluster.}

Although we were primarily concerned with interactions between BH-binaries and {\edit single background objects (stars and remnants)}, some situations required us to model encounters between single BHs and background binaries.  If the BH that we were following lost its companion as the result of a merger or a disruptive encounter, we used the method described in CSC13 to determine whether or not the BH would be able to exchange back into a binary.  {\edit This physical process depends upon} some assumptions about the nature and size of the binary populations present in GCs.  {\redit We considered three binary populations with different compositions.  One population, labeled OBS, was motivated by observational studies of photometric binaries in GCs, and was composed primarily of binaries that contained two main sequence stars \citep[e.g.,][]{Milone:2012}.  The second population, labeled FIR, was motivated by the theoretical study presented in \citet[][]{Fregeau:2009}.  In addition to showing that the hard binary fraction in the core of a cluster will increase over time, this work found that most binaries in the core will harbor at least one non-luminous member, making it difficult for observational studies to identify these systems.  The final binary population was optimized to interact with the single BHs and is labeled the OPT population.  None of the background binary populations included BHs.}  We varied the size of the binary population by adjusting the binary fraction, $f_{\rm b}$, within observational constraints. 

{\redit  A single BH could also be incorporated into a binary that is produced as the result of an encounter between the BH and two single stars.  The cross-section for such an encounter is typically very small, but it scales as roughly $M_{\rm BH}^5$.  We used the time-scale for three-body binary formation given in \citet{Ivanova:2005} to assess whether this processes is likely to occur before a single BH would exchange into an existing binary, as described above.  In most cases the exchange time-scale of a few Gyr is orders of magnitude shorter than the three-body binary formation time-scale (CSC13).  However, in simulations of cluster D that used a 35\msun BH, or several lower mass BHs, the two time-scales are within a factor of $\sim 2$ of each other.  Although three-body binary formation becomes competitive with exchanges in these extreme scenarios, we did not include this process in our models because it does not significantly reduce the amount of time it takes a BH to regain a binary companion.}       

\section{BH+MSP Binary Orbital Parameters}
\label{sec:orbparms}
The results of our simulations are summarized in \cref{tab:runs}.  The first column lists an identification number for each simulation.  The next six columns describe the initial conditions used for the simulations, noting the background cluster, binary population, binary fraction ($f_{\rm b}$), NS retention fraction ($f_{\rm ret}$), the BH mass ($M_{\rm BH}$), and the number of BHs ($N_{\rm BH}$).  The final six columns list the number of BH+NS binaries produced in that simulation ($N_{\rm BH+NS}$), {\edit the number of BH+BH binaries produced in the simulation ($N_{\rm BH+BH})$}, the median orbital period ($P_{\rm B}$) of the BH+NS binaries, {\edit the standard deviation of the BH+NS binary orbital period distribution ($s$)}, the fraction of time that a BH+NS binary exists ($\tau_{\rm BH+NS}$, {\redit see \cref{sec:detection}}), and {\edit the number of BH+NS binaries expected to be present in the cluster at any given time} ($p_{\rm BH+ NS}$, {\redit see \cref{sec:detection}}). We note that our code did not track whether or not a NS had been recycled into a MSP.  Thus, in the discussion that follows we will examine the nature of all of the BH+NS binaries produced in our simulations.  We will consider the fraction of NSs that are MSPs in \cref{sec:detection}.           

\begin{table*}
     \caption{GC Models and BH+MSP Binary Properties \label{tab:runs}}
     \begin{tabular}{rccrrrrrrrrrr}
	\hline
	Simulation & 
	Cluster & 
	Binary &
	$f_{\rm b}$ &
	$f_{\rm ret}$ &
	$M_{\rm BH}$ &
	$N_{\rm BH}$ & 
	$N_{\rm BH+NS}$&
	$N_{\rm BH+BH}$&
	Median $P_{\rm B}$ &
	$s$ &
	$\tau_{\rm BH+NS}$ &
	$p_{\rm BH+NS}$\\
	ID &
	Family &
	Population &
	 &
	 &
	($\rm{M_{\odot}}$) &
	 &
	 &
	 &
	(day) &
	(day) &
	 &
	 \\ \hline
	
   1 &    A &     FIR &    0.50 &    0.20 &    7 &      1 &    10 &  -- & 28000   &   3800 & $ 1.4\times 10^{-3} $ & $ 1.4\times 10^{-3} $\\
   2 &    A &     FIR &    0.50 &    0.10 &    7 &    192 &     1 & 41  & 58000   &   3200 & $ 6.8\times 10^{-5} $ & $ 1.3\times 10^{-2} $\\
   3 &    A &     FIR &    0.20 &    0.20 &   35 &      1 &    1 &  --  & 11000   &   2200 & $ 2.1\times 10^{-4} $ & $ 2.1\times10^{-4} $\\
   4 &    A &     FIR &    0.50 &    0.20 &   35 &      1 &     2 & --  & 14000   &   3300 & $ 3.3\times 10^{-4} $ & $ 3.3\times 10^{-4} $\\
   5 &    A &     FIR &    0.75 &    0.20 &   35 &      1 &     3 & --  & 2700    &   1600 & $ 1.3\times 10^{-3} $ & $ 1.3\times 10^{-3} $\\[1ex]
   6 &    A &     OPT &    0.20 &    0.20 &   35 &      1 &     7 & --  & 9500    &   9000 & $ 2.0\times 10^{-3} $ & $ 2.0\times 10^{-3} $\\[1ex]
   7 &    A &     OBS &    0.05 &    0.20 &   35 &      1 &     4 & --  & 12000   &  12000 & $ 1.2\times 10^{-3} $ & $ 1.2\times 10^{-3} $\\
   8 &    A &     OBS &    0.10 &    0.20 &   35 &      1 &    11 & -- & 13000   &  11000 & $ 1.8\times 10^{-3} $ & $ 1.8\times 10^{-3} $\\
   9 &    A &     OBS &    0.20 &    0.20 &   35 &      1 &     6 & -- & 2500    &  11000 & $ 1.3\times 10^{-3} $ & $ 1.3\times 10^{-3} $\\[2ex]
  10 &    B &     FIR &    0.50 &    0.20 &    7 &      1 &   466 & --  & 95 &    830 & $ 7.1\times 10^{-2} $ & $ 7.1\times 10^{-2} $\\
  11 &    B &     FIR &    0.50 &    0.20 &    7 &     42 &   183 & 290 & 260 &   1600 & $ 1.8\times 10^{-2} $ & $ 7.5\times 10^{-1} $\\
  12 &    B &     FIR &    0.50 &    0.10 &    7 &     46 &    84 & 361 &150 &   1100 & $ 1.1\times 10^{-2} $ & $ 5.3\times 10^{-1} $\\
  13 &    B &     FIR &    0.50 &    0.10 &    7 &     86 &    42 & 374 & 390 &   2411 & $ 3.8\times10^{-3} $ & $ 3.3\times10^{-1} $\\
  14 &    B &     FIR &    0.50 &    0.10 &   15 &      1 &   245 & --  & 47 &    500 & $ 3.3\times 10^{-2} $ & $ 3.3\times 10^{-2} $\\
  15 &    B &     FIR &    0.50 &    0.10 &   15 &      8 &    69 & 370 & 71 &    990 & $ 5.2\times 10^{-3} $ & $ 4.1\times 10^{-2} $\\
  16 &    B &     FIR &    0.50 &    0.10 &   15 &     17 &    37 & 391 & 92 &   1030 & $ 3.0\times 10^{-3} $ & $ 5.1\times 10^{-2} $\\
  17 &    B &     FIR &    0.20 &    0.20 &   35 &      1 &   327 &  -- & 16      &    340  & $ 4.3\times 10^{-2} $ & $ 4.3\times 10^{-2} $\\
  18 &    B &     FIR &    0.50 &    0.20 &   35 &      1 &   360 & --  & 25      &    360 & $ 4.4\times 10^{-2} $ & $ 4.4\times 10^{-2} $\\
  19 &    B &     FIR &    0.75 &    0.20 &   35 &      1 &   374 & --  & 22 &    400 & $ 4.3\times 10^{-2} $ & $ 4.3\times 10^{-2} $\\[1ex]
  20 &    B &     OPT &    0.20 &    0.20 &   35 &      1 &   364 & --  & 18 &    310 & $ 5.0\times 10^{-2} $ & $ 5.0\times 10^{-2} $\\[1ex]
  21 &    B &     OBS &    0.10 &    0.10 &    7 &     46 &    66 & 340 & 190 &   1500 & $ 1.0\times 10^{-2} $ & $ 4.6\times 10^{-1} $\\
  22 &    B &     OBS &    0.10 &    0.10 &   15 &      8 &    63 & 331 & 56 &    780 & $ 6.4\times 10^{-3} $ & $ 5.1\times 10^{-2} $\\
  23 &    B &     OBS &    0.05 &    0.20 &   35 &      1 &   341 & --  & 25 &    480 & $ 3.8\times 10^{-2} $ & $ 3.8\times 10^{-2} $\\
  24 &    B &     OBS &    0.10 &    0.20 &   35 &      1 &   320 & --  & 27 &    340 & $ 3.7\times 10^{-2} $ & $ 3.7\times 10^{-2} $\\
  25 &    B &     OBS &    0.20 &    0.20 &   35 &      1 &   337 & --  & 24 &    400 & $ 3.8\times 10^{-2} $ & $ 3.8\times 10^{-2} $\\[2ex]
  26 &    C &     FIR &    0.50 &    0.20 &    7 &      1 &  2524 & --  & 6.7 &    130 & $ 8.4\times 10^{-2} $ & $ 8.4\times 10^{-2} $\\
  27 &    C &     FIR &    0.50 &    0.10 &    7 &     14 &  1191 & 454  & 8.5 &    190 & $ 4.6\times 10^{-2} $ & $ 6.5\times 10^{-1} $\\
  28 &    C &     FIR &    0.50 &    0.20 &    7 &     14 &  2061 & 355 & 8.8 &    210 & $ 7.6\times 10^{-2} $ & $ 1.1 $\\
  29 &    C &     FIR &    0.50 &    0.05 &    7 &     14 &   839 & 552 & 7.8 &    190 & $ 3.3\times 10^{-2} $ & $ 4.7\times 10^{-1} $\\
  30 &    C &     FIR &    0.50 &    0.10 &   15 &      1 &  2146 & --  & 5.6 &    150 & $ 4.8\times 10^{-2} $ & $ 4.8\times 10^{-2} $\\
  31 &    C &     FIR &    0.50 &    0.10 &   15 &      9 &   333 & 955 & 9.5 &    250 & $ 6.3\times 10^{-3} $ & $ 5.7\times 10^{-2} $\\
  32 &    C &     FIR &    0.50 &    0.10 &   15 &     19 &   167 & 956 & 26 &    390 & $ 1.7\times 10^{-3} $ & $ 3.2\times 10^{-2} $\\
  33 &    C &     FIR &    0.20 &    0.20 &   35 &      1 &  1209 & --  & 4.8 &     76 & $ 2.2\times 10^{-2} $ & $ 2.2\times 10^{-2} $\\
  34 &    C &     FIR &    0.50 &    0.20 &   35 &      1 &  1868 & --  & 4.8 &     89 & $ 3.2\times 10^{-2} $ & $ 3.2\times 10^{-2} $\\
  35 &    C &     FIR &    0.75 &    0.20 &   35 &      1 &  3042 & --  & 4.8 &     92 & $ 5.6\times 10^{-2} $ & $ 5.6\times 10^{-2} $\\[1ex]
  36 &    C &     OPT &    0.20 &    0.20 &   35 &      1 &  2071 & --  & 5.1 &    100 & $ 3.5\times 10^{-2} $ & $ 3.5\times 10^{-2} $\\[1ex]
  37 &    C &     OBS &    0.10 &    0.10 &    7 &     14 &   927 & 318 & 8.8 &    220 & $ 3.4\times 10^{-2} $ & $ 4.7\times 10^{-1} $\\
  38 &    C &     OBS &    0.10 &    0.10 &   15 &      9 &   245 & 710 & 12 &     280 & $ 4.1\times 10^{-3} $ & $ 3.7\times 10^{-2} $\\
  39 &    C &     OBS &    0.05 &    0.20 &   35 &      1 &  1148 & --  & 4.5 &     80 & $ 1.8\times 10^{-2} $ & $ 1.8\times 10^{-2} $\\
  40 &    C &     OBS &    0.10 &    0.20 &   35 &      1 &  1081 & --  & 4.6 &    100 & $ 1.8\times 10^{-2} $ & $ 1.8\times 10^{-2} $\\
  41 &    C &     OBS &    0.20 &    0.20 &   35 &      1 &  1304 & --  & 4.5 &     90 & $ 2.2\times 10^{-2} $ & $ 2.2\times 10^{-2} $\\[2ex]
  42 &    D &     FIR &    0.50 &    0.20 &    7 &      1 &  4236 & --  & 4.6 &    160 & $ 4.4\times 10^{-2} $ & $ 4.4\times 10^{-2} $\\  
  43 &    D &     FIR &    0.10 &    0.10 &   15 &     18 &   190 & 779 & 4.6 &    270 & $ 8.9\times 10^{-4} $ & $ 1.6\times 10^{-2} $\\
  44 &    D &     FIR &    0.20 &    0.20 &   35 &      1 &  2676 & --  & 3.2 &    120 & $ 1.5\times 10^{-2} $ & $ 1.5\times 10^{-2} $\\
  45 &    D &     FIR &    0.50 &    0.20 &   35 &      1 &  5579 & --  & 3.2 &    120 & $ 3.3\times 10^{-2} $ & $ 3.3\times 10^{-2} $\\
  46 &    D &     FIR &    0.75 &    0.20 &   35 &      1 &  8524 & --  & 3.2 &    110 & $ 4.9\times 10^{-2} $ & $ 4.9\times 10^{-2} $\\[1ex]
  47 &    D &     OPT &    0.20 &    0.20 &   35 &      1 &  8212 & --  & 3.8 &    140 & $ 4.2\times 10^{-2} $ & $ 4.2\times 10^{-2} $\\[1ex]
  48 &    D &     OBS &    0.10 &    0.10 &   15 &     18 &   260 & 914 & 6.5 &    290 & $ 1.3\times 10^{-3} $ & $ 2.3\times 10^{-2} $\\
  49 &    D &     OBS &    0.05 &    0.20 &   35 &      1 &  1753 & --  & 3.4 &    110 & $ 9.2\times 10^{-3} $ & $ 9.2\times 10^{-3} $\\
  50 &    D &     OBS &    0.10 &    0.20 &   35 &      1 &  2029 & --  & 3.3 &    120 & $ 1.1\times 10^{-2} $ & $ 1.1\times 10^{-2} $\\
  51 &    D &     OBS &    0.20 &    0.20 &   35 &      1 &  2939 & --  & 3.5 &    130 & $ 1.6\times 10^{-2} $ & $ 1.6\times 10^{-2} $\\
\end{tabular}
\end{table*}
\begin{figure*}
	\centering
	\includegraphics[width=1.0\textwidth]{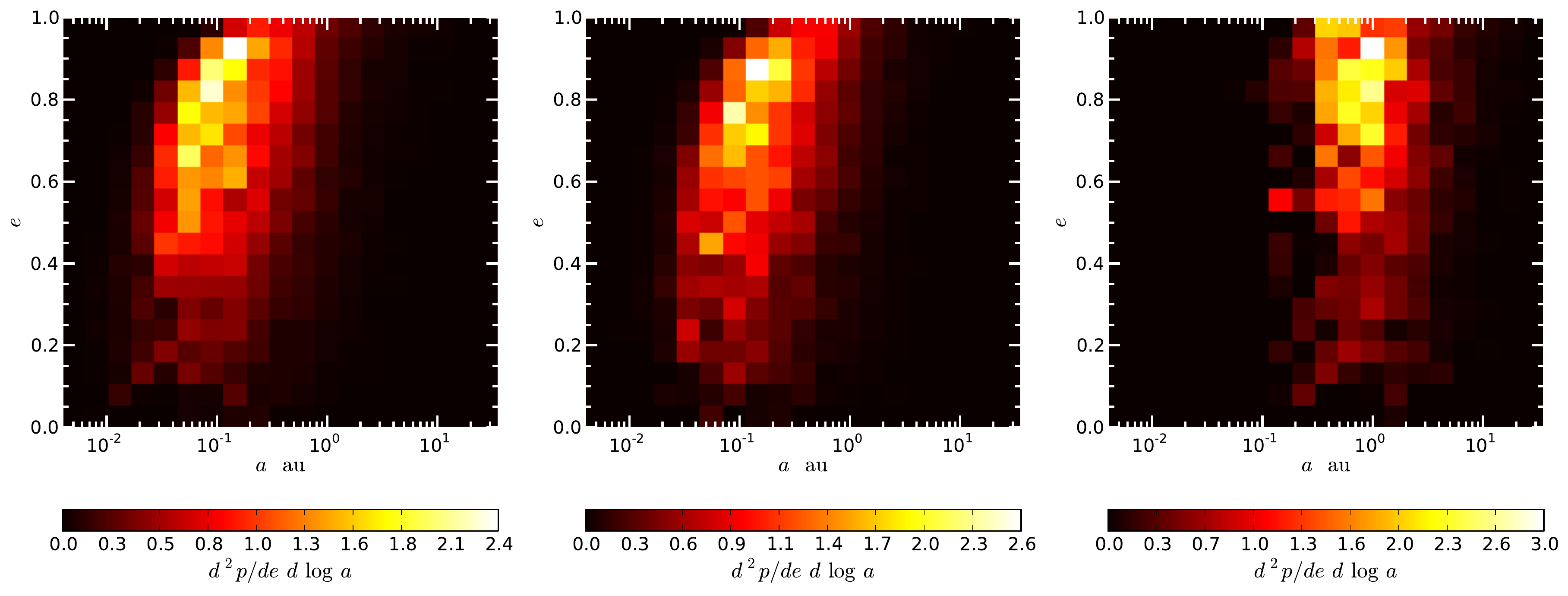}
	\caption{Joint $e-a$ distribution for BH+NS binaries produced in simulations 42 (left panel), 26 (center panel), and 10 (right panel).  The eccentricity distribution is similar in all of these simulations.  The semimajor axis distribution shifts to larger $a$ as the cluster central density decreases from $n_{\rm c} = 10^{6}$ in the left panel to $n_{\rm c} = 10^{5}$ in the right panel.  \label{fig:ae}}
\end{figure*}

{\edit  Encounters between BH-binaries that do not contain a NS and background main sequence stars, giants, or WDs can influence the nature of the BH+NS binary population.  These interactions occur more frequently because these objects are far more common than NSs.  Such encounters can limit the size of the BH+NS binary population by removing BHs from the cluster as a result of a super-elastic binary--single collision.  However, encounters of this strength were rare in our simulations.  In most cases, more that 75\% of the 2000 BHs that we followed during a simulation were retained by the cluster for the entire $10^{10}$ yr run.  In the rare cases where most of the BHs were ejected, the majority of BHs were ejected after merging with a NS and receiving a recoil that exceeded the cluster's escape velocity {\redit \citep{Shibata:2009}}.  While encounters with non-NSs did not efficiently remove BHs from the cluster, these interactions did impact the size of the BH+NS star population by destroying BH-binaries.  An interaction can either disrupt the binary into three single stars or induce {\redit a physical collision} and merger between the BH and its companion.  Either way, at the end of the encounter the BH has lost its companion and must begin the slow process of exchanging back into a binary, delaying the formation of a BH+NS binary.  We defer the discussion of these encounters and BH-binaries, which are potential X-ray sources, to a future publication.  Below we will focus on the properties of the BH+NS binary population and encounters involving these binaries.}    

In our simulations, there were two processes that drove the evolution of a BH+NS binary's orbital parameters.  Encounters between the binary and background stars changed the semimajor axis ($a$) and eccentricity ($e$) of the binary impulsively.  In most cases, an encounter resulted in the binary becoming more tightly bound or ``hardened.''  The emission of gravitational radiation also modified the BH+NS binaries's orbital parameters, driving $a$ and $e$ towards zero.  Since the orbital parameters of the binaries were constantly changing, we resampled the simulation output in even time intervals to ensure that each orbital configuration that the BH+NS binaries evolved through was properly weighted.  We chose a resampling time step of $10^{7}$ yr.  We checked that our choice of $10^{7}$ yr intervals did not bias the resampled orbital parameter distributions by repeating the analysis of a subset of our simulations with finer ($10^{5}$ yr) time resolution.  A coarser resampling, with steps of $10^{8}$ yr, failed to capture the wings of the semimajor axis distribution, where rapid evolution occurs.  Examples of the resampled data from three representative simulations are shown in \cref{fig:ae}.  The figure shows the joint distribution for the eccentricity and semimajor axis of the BH+NS binaries produced in simulations using background clusters B, C, and D.           

\subsection{Eccentricity distribution}
\label{sec:ecc}
{\edit In nearly all of our simulations} the eccentricity distribution of the BH+NS binaries was roughly thermal for $e \la 0.9$. At higher eccentricity the distribution flattened out and turned over.  That is to say, there were fewer binaries with $e \ga 0.9$ than expected given the $f(e) = 2e$ distribution.  This is because the emission of gravitational radiation acted to quickly circularize such high eccentricity systems.  Given these results, we expect that the mean $e$ of dynamically formed BH+NS binaries in GCs will be in the range 0.6--0.7.  {\edit In the skewed eccentricity distributions seen in our simulations, the median value of $e$ exceeds the mean so most binaries will have eccentricities larger than these average values}.  The only simulations that did not result in a thermal eccentricity distribution were those run in background cluster A.  As can be seen in \cref{tab:runs}, very few BH+NS binaries were produced in this low density cluster.  We were, therefore, unable to study the eccentricity distributions in these poorly sampled cases.       

\subsection{Orbital separation distribution}
Of all of the parameters varied in our study, the background cluster's structural properties had the strongest impact on the semimajor axes of the BH+NS binaries.  For this reason, we will describe the distributions of semimajor axes and orbital periods cluster-by-cluster.  Once we have described the results for each cluster, we will investigate the origin of the observed trends.  

GC A was the least massive cluster considered in our study.  Furthermore, because of its low density, this cluster also had the lowest encounter rate.  NSs only exchanged into 45 of the $1.8\times10^{4}$ BH-binaries that we evolved in this cluster.  {\edit The corresponding specific BH+NS binary formation rate of $\sim 3 \times 10^{-10}$ yr$^{-1}$ \msun$^{-1}$ is  smaller than the specific BH+recycled pulsar formation rate predicted for the field.  Although the formation rate of BH+MSP binaries is not enhanced by dynamical processes in this low density cluster, given the short lifetimes of BH+recycled pulsar binaries formed through standard evolutionary channels (see \cref{sec:intro}), it is likely that any BH+MSP binaries present in such a cluster today were formed through dynamical interactions. Thus, the properties of these clusters' present day BH+NS binary populations should be described by the systems formed in our simulations.} The semimajor axes of these rare BH+NS binaries spanned a range of nearly two orders of magnitude, from 1.2--113 au.  The BH+NS binaries produced in this cluster were extremely wide because neither of the hardening mechanisms described above were effective.  The binaries underwent very few encounters because of the cluster's low density, and the effects of gravitational radiation were negligible at such large orbital separations.  Accordingly, we predict that any BH+MSP binaries present in low density clusters are likely to have orbital periods of several decades.

\begin{figure*}
	\centering
	\includegraphics[width=0.85\textwidth]{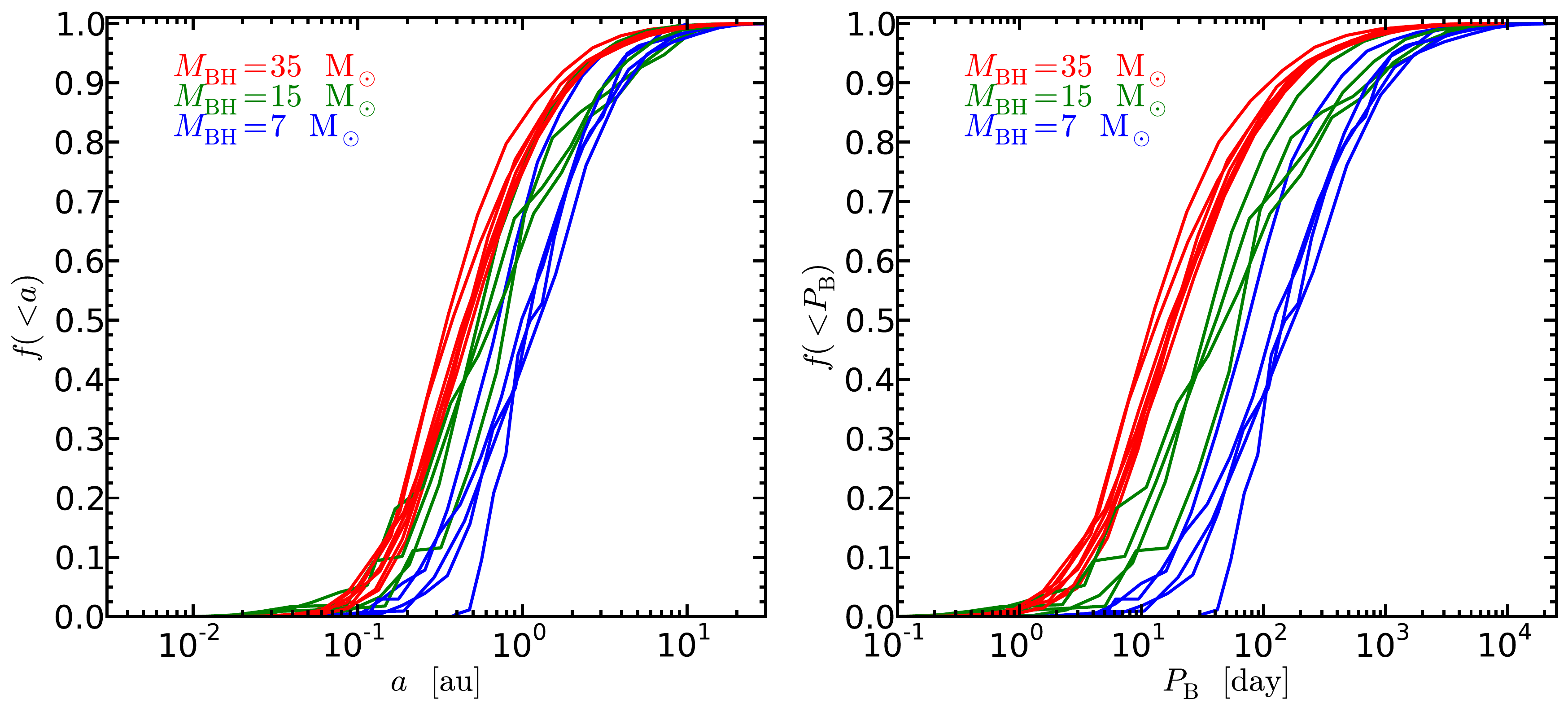}
	\caption{Cumulative distributions of the semimajor axis (left) and orbital period (right) for a subset of the BH+NS binaries produced in each of the simulations in cluster B.  Each curve shows the distribution of $a$ or $P_{\rm B}$ in a particular simulation.  The distributions shown only include systems that exist at $t > 8 \times 10^{9}$ yr, so they correspond to the present day.  The color of each curve denotes the mass of the BH(s) used in a particular simulation, with 35 ${\rm M_{\sun}}$, 15 ${\rm M_{\sun}}$, and 7 ${\rm M_{\sun}}$ BHs denoted by red, green, and blue, respectively. Note the clear relationship between $M_{\rm BH}$ and orbital separation: simulations with more massive BHs produced BH+NS binaries with {\em smaller} semimajor axes. This is the opposite of the trend seen in higher density clusters. \label{fig:bdists}}
\end{figure*}

The dynamically formed BH+NS binaries in GC B had significantly smaller orbital separations.  In this higher density cluster, frequent encounters hardened some of the BH+NS binaries to small enough $a$ that gravitational radiation effects became important.  \cref{fig:bdists} shows the cumulative distributions of $a$ and $P_{\rm B}$ amongst present day BH+NS binaries for all of our simulations in cluster B (i.e., simulations 10-25).  We constructed a present day population by only selecting binaries that exist at $t > 8\times10^{9}$ yr.  The median values of $a$ in these distributions fell in the range 0.42--1.6 au.  It is evident from  \cref{fig:bdists} that the orbital separation is influenced by the value of $M_{\rm BH}$.  Simulations with higher mass BHs produced tighter BH+NS binaries.  This trend is amplified in the orbital period distributions because the semimajor axis and orbital period are linked by an additional factor of $\sim M_{\rm BH}^{-1/2}$, specifically $P_{\rm B} = \sqrt{a^{3}/(M_{\rm BH} + M_{\rm NS})}$.  At a given value of $M_{\rm BH}$, the orbital period distributions are fairly similar to one another, despite significantly different assumptions about the number of BHs in the clusters and the clusters' binary populations.  This suggests that these properties do not strongly impact the orbital periods of the BH+NS binaries produced in GCs with structure similar to that of cluster B.  The median BH+NS binary orbital periods in all of the simulations in cluster B fell between 16 days and 260 days.  The standard deviation of the $P_{\rm B}$ distribution within a particular simulation was much wider (see \cref{tab:runs}).  

\begin{figure*}
	\centering
	\includegraphics[width=0.85\textwidth]{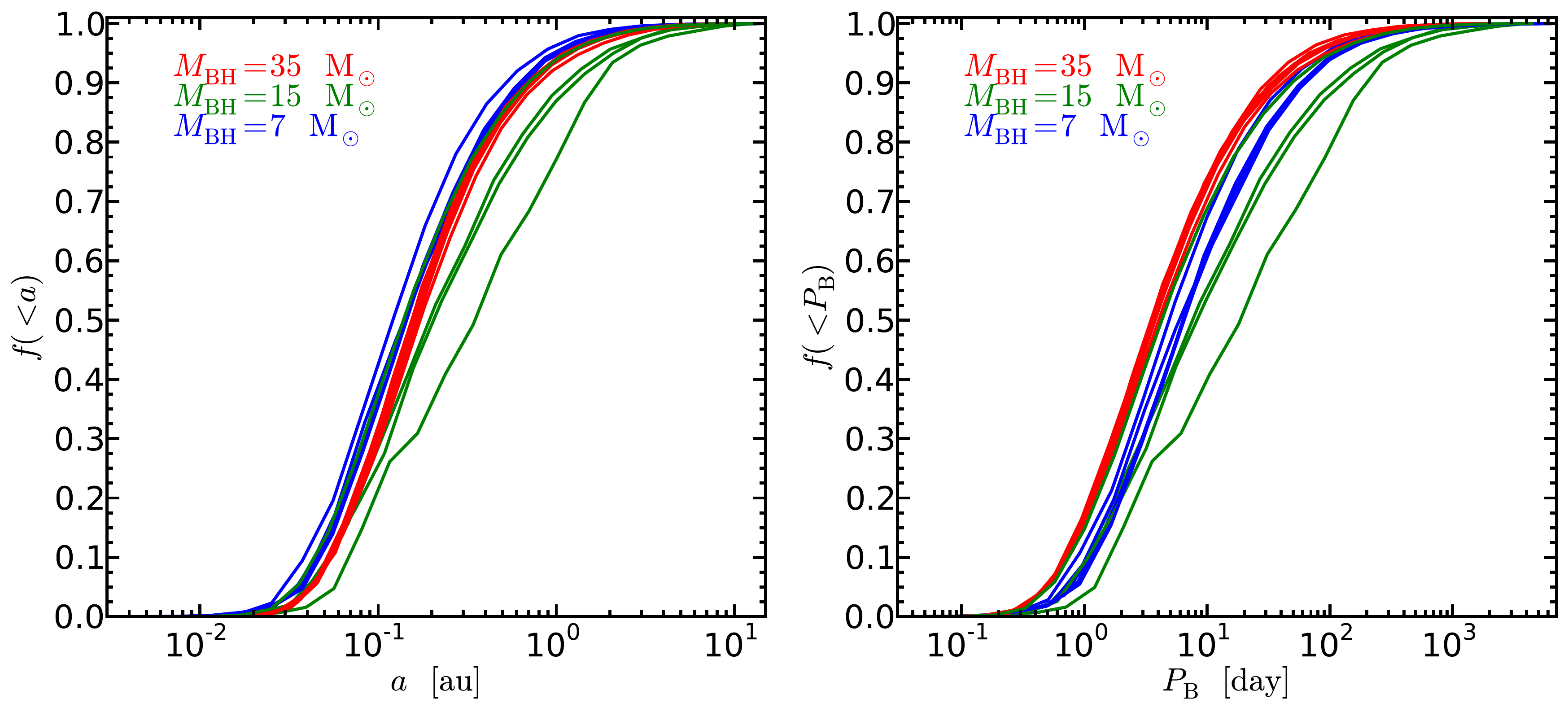}
	\caption{Cumulative distributions of the semimajor axis (left) and orbital period (right) for a subset of the BH+NS binaries produced in each of the simulations in cluster C.  Each curve shows the distribution of $a$ or $P_{\rm B}$ in a particular simulation.  The distributions shown only include systems that exist at $t > 8 \times 10^{9}$ yr, so they correspond to the present day.  The color of each curve denotes the mass of the BH(s) used in a particular simulation, with 35 ${\rm M_{\sun}}$, 15 ${\rm M_{\sun}}$, and 7 ${\rm M_{\sun}}$ BHs denoted by red, green, and blue, respectively. Note the clear relationship between $M_{\rm BH}$ and orbital separation: simulations with more massive BHs produce BH+NS binaries with {\em larger} semimajor axes. This is contrary to the trend seen in the lower density cluster B. Note, however, that simulations with $M_{\rm BH} = 15\msun$ \emph{and} $N_{\rm BH} >1$ do not follow this trend.   \label{fig:cdists}}
\end{figure*}

For simulations in cluster C, the BH+NS binary orbital separations were smaller still.  The cumulative distributions of the semimajor axis and orbital periods for the BH+NS binaries formed in this cluster are shown in \cref{fig:cdists}.  The median values of $a$ in these simulations ranged between 0.15 au and 0.43 au.  Simulation 32 resulted in the largest median $a$, and from \cref{fig:cdists} it is clear that the cumulative $a$ and $P_{\rm B}$ distributions for simulation 32 deviate from distributions seen in other simulations.  The critical difference between this simulation and the others run in cluster C was the large number of BHs $(N_{\rm BH} = 19)$ retained by the cluster.  {\edit As we will discuss in \cref{sec:detection}, the evolution of the BH+NS binaries can be significantly altered when there are several relatively massive BHs present in the cluster.  If we exclude the simulations that used more than one 15 \msun BH (i.e., simulations  31, 32, and 38) from our analysis, the range of median semimajor axes reduces to 0.15--0.20 au.}  This range is much narrower than that observed in our simulations in cluster B.  The semimajor axes of the BH+NS binaries produced in cluster C only depended weakly on many of the input parameters, including $M_{\rm BH}$ and $f_{\rm b}$.  In addition to being a much smaller effect, the relationship between BH mass and semi-axis seen in cluster C is the reverse of what was seen in cluster B.  Here, the simulations with the lowest mass BHs produced the BH+NS binaries with the smallest semimajor axes.  Most of the BH+MSP binaries in GCs with structures similar to cluster C will have orbital periods shorter than 10 days.        

\begin{figure*}
	\centering
	\includegraphics[width=0.85\textwidth]{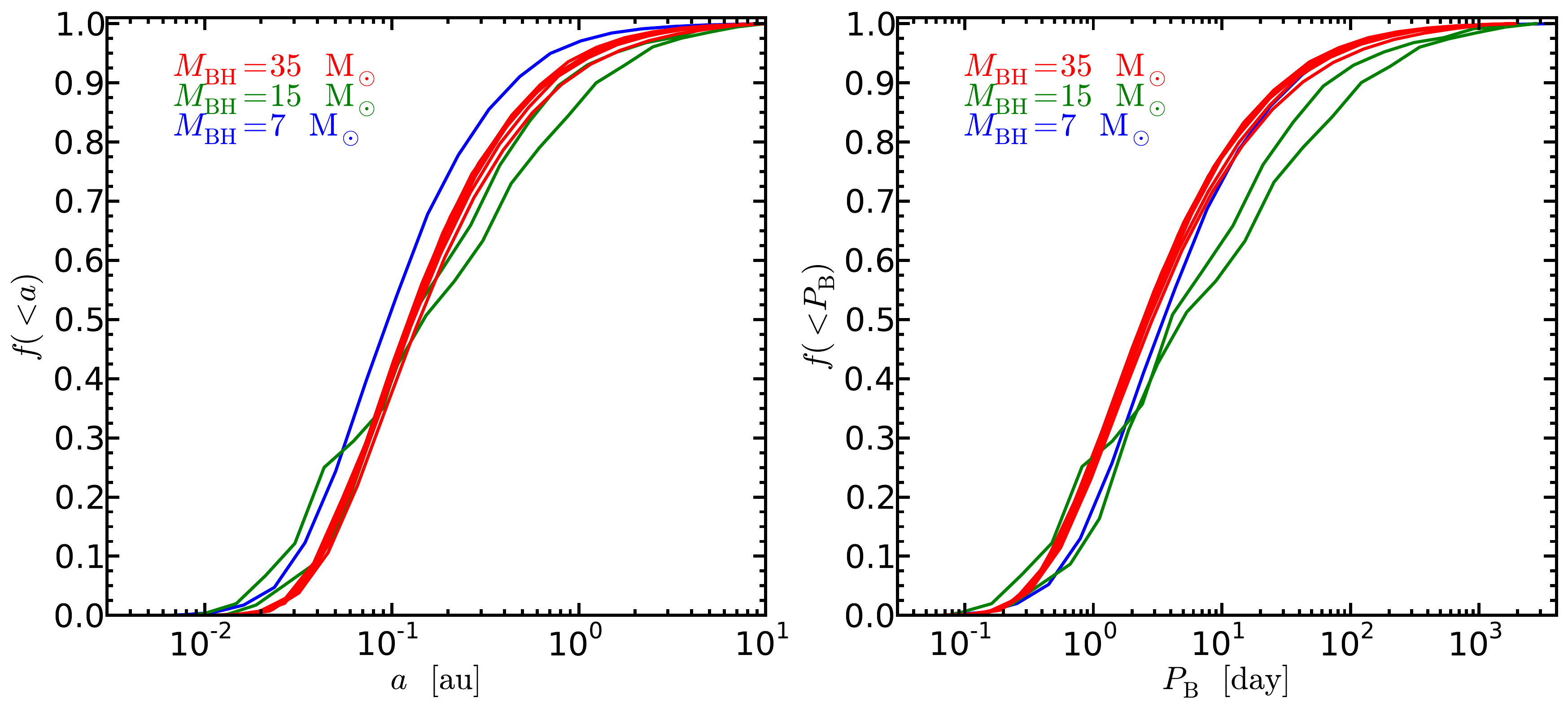}
	\caption{Cumulative distribution of the semimajor axis (left) and orbital period (right) for a subset of the BH+NS binaries produced in each of the simulations in cluster family D.  Each curve shows the distribution of $a$ or $P_{\rm B}$ in a particular simulation.  The distributions shown only include systems that exist at $t > 8 \times 10^{9}$ yr so they correspond to the present day.  Note that simulations with 7 ${\rm M_{\odot}}$ BHs produce binaries with smaller semimajor axes than the simulations that used 35 ${\rm M_{\odot}}$ BHs.  This is contrary to the trend seen in the lower density cluster B.  \label{fig:ddists}}
\end{figure*}

Finally, the shortest period BH+NS binaries observed in our simulations were produced in cluster D.  The encounter rate was highest in cluster D, so BH-binaries were rapidly hardened to small orbital separations in the simulations that used this background cluster.  \cref{fig:ddists} shows the present day cumulative distributions of $a$ and $P_{\rm B}$ for each of the simulations performed in cluster D  (i.e., simulations 42-51).  As was seen in the simulations that used cluster C, the simulation with $M_{\rm BH} = 7~{\rm M_{\odot}}$ produced BH+NS binaries with the smallest semimajor axes.  However, the median value of $a$ only varied slightly from simulation to simulation in cluster D.  Simulation 42 had the smallest median $a = 0.11$ au.  The median value of $a$ was largest in simulations 47 and 48, both of which had median $a = 0.17$ au.  We expect that many BH+MSP binaries present in densest GCs will have $P_{\rm B} \la 5$ days.          

To review, we have shown that the BH+NS binaries produced in high density clusters have smaller semimajor axes than those produced in low density clusters.  Furthermore, in the high density GCs, the binaries with lower mass BHs had smaller orbital separations.  However, the size of a BH+NS binary's semimajor axis only depended weakly on the mass of the BH.  When we changed $M_{\rm BH}$ by a factor of 5, the median semimajor axis only changed by a factor of $\sim 1.5$. The opposite trend was seen in lower density clusters (e.g., cluster B). In such clusters, binaries with higher mass BHs tended to have smaller semimajor axes than those with low mass BHs.  Additionally, the mass of the BHs had a stronger impact on the orbital separations of the binaries produced in these simulations (compare \cref{fig:bdists} and \cref{fig:ddists}).  It is clear that the BH+NS binaries present in high density clusters  (similar to clusters C and D) are in a different evolutionary phase than those in lower density clusters (similar to cluster B).   

This behavior can be understood by comparing the rates of the two evolutionary processes described above: encounters with single stars and the emission of gravitational radiation.  For hard binaries, encounters with background stars, on average, increase the binding energy of the binary at a constant rate
\begin{equation}
	\left<\frac{dE_{\rm B}}{dt}\right> = A_{\rm bs} \frac{G^{2}  M_{\rm BH}  M_{\rm NS} (M_{\rm BH} + M_{\rm NS})n_{\rm c}} {\bar{v}_{\rm m}},
	\label{eqn:dedtenc}
\end{equation}
where $G$ is the gravitational constant and $A_{\rm bs}$ is a dimensionless parameter of order unity \citep{Heggie:1975}.  Here we have used the central values of the stellar density and velocity dispersion, $n_{\rm c}$ and $\bar{v}_{\rm m}$. The BHs were the most massive members of the GCs that we modeled, so they spent most of their lifetimes deep in the cluster cores. Thus, using the central values will give a reasonable approximation of the encounter rate.  This constant hardening rate implies that encounters change the semimajor axes of the binaries at a rate
\begin{equation}
	\left<\frac{da}{dt}\right>_{\rm enc} \sim \frac{-2 G (M_{\rm BH} + M_{\rm NS}) n_{c} a^{2}}{\bar{v}_{\rm m}}.
	\label{eqn:dadtenc}
\end{equation}
\citet{Peters:1964} gives the rate at which gravitational radiation shrinks a BH+NS binary's orbit as:
\begin{equation}
  \begin{split}
		\label{eqn:dadtgw}                                                               
	\left<\frac{da}{dt}\right>_{\rm GW} = -\frac{64}{5} \frac{G^{3} M_{\rm BH} M_{\rm NS}  (M_{\rm BH} + M_{\rm NS})}{c^{5} a^{3} (1-e^{2})^{7/2}}\\  \times \left(1+\frac{73}{24}e^{2}+\frac{37}{96}e^{4}\right)   
 \end{split}
\end{equation}
where $c$ is the speed of light.   

Now let us compare these rates.  For wide binaries, $|\langle da/dt \rangle_{\rm enc}| \gg |\langle da/dt\rangle_{\rm GW}|$ and encounters with single stars will be the dominant hardening mechanism.   On the other hand, the evolution of binaries with small orbital separations will occur more rapidly through the emission of gravitational radiation than through encounters with background stars.  The initial semimajor axes of the dynamically formed BH+NS binaries considered here will be relatively large, as is typical of systems produced in exchange interactions.  Thus, the semimajor axis of a newly formed BH+NS binary will shrink fairly rapidly as a result of encounters with single stars. As encounters reduce the binary's orbital separation, $|\langle da/dt \rangle_{\rm enc}|$ also declines. Eventually, $a$ (and $|\langle da/dt \rangle_{\rm enc}|$) will be such that encounters do not efficiently harden the binary, and the orbital evolution will effectively stall.  The binary will spend a large fraction of its lifetime at roughly constant semimajor axis, awaiting a final encounter that will reduce its semimajor axis to the point at which gravitational radiation takes over.  At this point the  binary's orbital evolution will occur more rapidly as $a$ continues to shrink.  Therefore, the semimajor axis distributions will be dominated by BH+NS binaries with orbital configurations that fall in the transition between the encounter and gravitational radiation dominated regimes.    

\begin{figure*}
	\centering
	\includegraphics[width=1.0\textwidth]{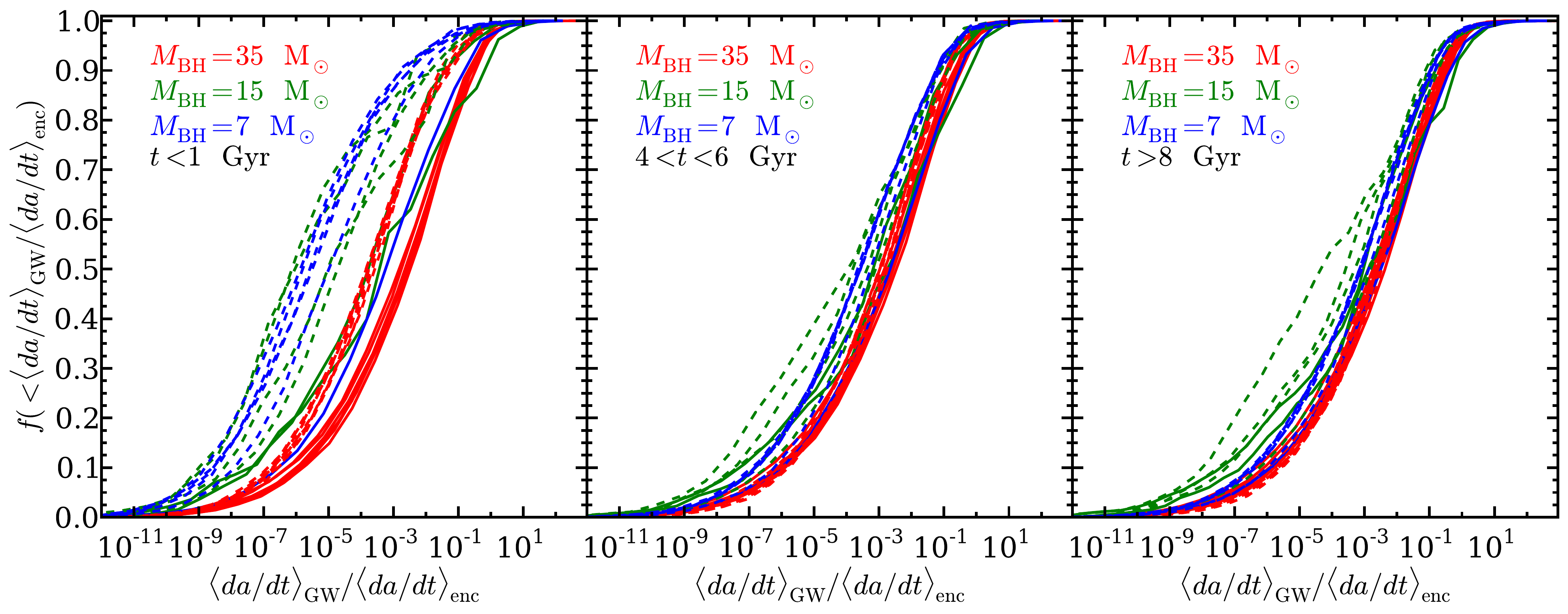}
	\caption{Cumulative distributions of the ratio of the rate of semimajor axis evolution due to the emission of gravitational radiation to rate of semimajor axis evolution due to encounters with background stars (given in \cref{eqn:dadtgw,,eqn:dadtenc}, respectively).   Each curve shows the distribution of $\langle da/dt \rangle_{\rm GW}/\langle da/dt\rangle_{\rm enc}$ in a particular simulation in cluster C (dashed lines) or cluster D (solid lines).  The three panels illustrate how the distribution evolves over time.  The distributions shown are for  $t< 1 \times 10^{9}$ yr (left panel), $4 \times 10^{9} < t < 6 \times 10^{9}$ yr (center panel), and  $t > 8 \times 10^{9}$ yr (right panel).  The color of each curve denotes the mass of the BH(s) used in a particular simulation, with 35 ${\rm M_{\sun}}$, 15 ${\rm M_{\sun}}$, and 7 ${\rm M_{\sun}}$ BHs denoted by red, green, and blue, respectively.  The binaries evolve towards a steady distribution in $\langle da/dt \rangle_{\rm GW}/\langle da/dt\rangle_{\rm enc}$.  Simulations with the largest encounter rates (those in cluster D and those in cluster C with $M_{\rm BH} = 35~{\rm M_{\odot}}$) have reached the steady configuration by the current epoch.  The simulations with lower encounter rates are still approaching this configuration, from the left, but have not reached this state after $10^{10}$ yr of evolution.  \label{fig:tscrat}}
\end{figure*}

We explored the transition between the encounter dominated phase and the gravitational radiation dominated phase by examining the ratio $\langle da/dt \rangle_{\rm GW}/\langle da/dt\rangle_{\rm enc}$.   Using \cref{eqn:dadtenc,eqn:dadtgw}, we calculated the ratio for each point in the BH+NS binaries' resampled evolutionary tracks. The cumulative distributions of $\langle da/dt \rangle_{\rm GW}/\langle da/dt\rangle_{\rm enc}$ from a subset of our simulations are shown in \cref{fig:tscrat}.  Here we have only plotted simulations in clusters C and D.  As discussed above, the semimajor axis distributions in these clusters exhibit similar trends. The left panel shows how $\langle da/dt \rangle_{\rm GW}/\langle da/dt\rangle_{\rm enc}$ is distributed at $t < 1 \times 10^{9}$ yr.   The center panel shows how these distributions have changed after $\sim 5$ Gyr of evolution.  Note that the distribution of $\langle da/dt \rangle_{\rm GW}/\langle da/dt\rangle_{\rm enc}$ for simulations of cluster D that used $35\msun$ BHs had hardly changed.  The distributions in all of the other simulations, with lower encounter rates, evolved towards these stationary curves.  The right panel shows how $\langle da/dt \rangle_{\rm GW}/\langle da/dt\rangle_{\rm enc}$ is distributed at the present day, after an additional $\sim 3$ Gyr of evolution.  Most of the $\langle da/dt \rangle_{\rm GW}/\langle da/dt\rangle_{\rm enc}$ distributions from simulations in cluster D (solid lines) lie on top of one another.  Simulations in cluster C (dashed lines) that used $M_{\rm BH} = 35~{\rm M_{\odot}}$ also closely follow this trend, while some of the simulations with lower mass BHs exhibit smaller values of $\langle da/dt \rangle_{\rm GW}/\langle da/dt\rangle_{\rm enc}$.    

We interpret this evolution as follows.  Given sufficient time, the distribution of $\langle da/dt \rangle_{\rm GW}/\langle da/dt\rangle_{\rm enc}$ in any cluster will approach the steady configuration seen in the right panel of \cref{fig:tscrat}.  At this stage, the orbital evolution of most of the BH+NS binaries has slowed as they make the transition from the encounter dominated phase to the gravitational radiation dominated phase.  The median value of $\langle da/dt \rangle_{\rm GW}/\langle da/dt\rangle_{\rm enc}$ in this steady configuration is $\sim 2\times10^{-3}$.  Even though $|\langle da/dt \rangle_{\rm GW}|$ is still a few orders of magnitude smaller than $|\langle da/dt \rangle_{\rm enc}|$ at this point, this ratio corresponds to the beginning of transition between the encounter and gravitational wave dominated phases.  Given the steep dependence of $\langle da/dt \rangle_{\rm GW}$ on $a$, such a binary is only about one encounter away from an orbit in which  $\langle da/dt \rangle_{\rm GW} \sim \langle da/dt\rangle_{\rm enc}$.  We note that despite the similarities illustrated in \cref{fig:tscrat}, the distributions from each simulation are formally distinct from one another, in most cases.  We performed the two-sample Kolmogorov-Smirnov test between each pair of distributions and found that only those from simulations 44, 45, and 46 were consistent with coming from the same parent distribution.  However, our interpretation does not require that the distributions be identical, it only requires that each distribution is dominated by systems that are in the transition phase.  

Since the orbital parameter distributions evolve towards a constant configuration that is dominated by binaries with $\langle da/dt \rangle_{\rm GW}\sim \langle da/dt\rangle_{\rm enc}$, we can use \cref{eqn:dadtenc,eqn:dadtgw} and derive a scaling relation for the median semimajor axis of the BH+NS binaries during this phase:
\begin{equation}
a \propto \left(\frac{M_{\rm BH} M_{\rm NS} \bar{v}_{\rm m}}{n_{\rm c}}\right)^{1/5}.
\label{eqn:am}
\end{equation}
In deriving this expression we have used the fact that the eccentricity distributions in every simulation were similar (see \cref{sec:ecc}). Many facets of this scaling relation are seen in the semimajor axis distributions from simulations computed in clusters C and D.  Most importantly this scaling relation accounts for the rather weak dependence of the these semimajor axis distributions on many of the input parameters. The relation also accounts for the fact that the simulations with lower mass BHs produced BH+NS binaries with smaller semimajor axes. 

Notably, many of the simulations that used $15\msun$ BHs do not seem to follow the evolutionary path sketched above.  \cref{fig:tscrat} shows that the BH+NS binaries produced in these simulations do not settle into the steady $\langle da/dt \rangle_{\rm GW}/\langle da/dt\rangle_{\rm enc}$ distribution seen in simulations that used both higher and lower mass BHs.  Furthermore, the semimajor axes of the BH+NS binaries produced in most of the simulations that used 15\msun BHs do not appear to follow the monotonic relationship with $M_{\rm BH}$ implied by \cref{eqn:am}.  If the BH+NS binaries did follow this relationship, one would expect the curves corresponding to simulations with 15\msun BHs (shown in green) to fall between the curves for simulations that used 7\msun (shown in blue) and 35\msun (shown in red) BHs in the left panels of \cref{fig:cdists,fig:ddists}.  The only simulation that obeys this relation is Simulation 30, which considered a cluster that only contained one 15\msun BH.  This suggests that the larger BH populations used in all of the other simulations with 15\msun BHs are likely to blame for the modified orbital evolution of the BH+NS in these models.  In fact, as we will discuss in \cref{sec:detection}, many of the BH+NS binaries produced in background clusters that contained a large population of fairly massive BHs were not hardened, but rather quickly destroyed when a second BH exchanged into the binary and ejected the NS.  This accounts for the lack of BH+NS binaries with small $a$ in the simulations that used 15\msun BHs.          

In clusters with low encounter rates, many of the BH+NS binaries will not be hardened fast enough to reach the transition from the encounter dominated regime to the gravitational radiation dominated regime within a Hubble time.  Therefore, the orbital parameters of most BH+NS binaries in these clusters are determined by binary--single encounters alone.  Accordingly, we expect that the semimajor axes of the binaries in these clusters will follow a different scaling relationship than those in the high encounter rate clusters explored above.  The rate at which encounters harden binaries is proportional to the mass of the BH (see \cref{eqn:dadtenc}), so we would expect for the simulations that used higher mass BHs to produce BH+NS binaries with smaller semimajor axes in the encounter dominated regime.  This is what we observed in simulations in cluster B; the binaries with higher mass BHs were hardened to smaller $a$ than the binaries with low mass BHs.  

\section{BH+MSP Binary Population Size}
\label{sec:detection}

The 51 simulations presented here investigated how several parameters impact the dynamical formation of BH+NS binaries in GCs.  Below we will describe how each of these traits affects the likelihood that a BH+NS binary exists within a cluster.  We will use two metrics to characterize the likelihood that a cluster harbors a BH+NS binary.  {\edit The first metric involves the average time fraction that a BH in our simulations had a NS companion, $\tau_{\rm{BH+NS}} = \sum_{i} t_{\rm BH + NS,i}/(N_{\rm runs} t_{\rm max})$.  Here $t_{\rm BH + NS,i}$ is the amount of time that a BH had a NS companion during a particular run, the sum is over all of the runs in a single simulation, $t_{\rm max} = 10^{10}$ yr is the duration of each run, and $N_{\rm runs}$ is the number of runs in the simulation.}  In practice, $N_{\rm runs}$ is slightly less than 2000 because we rejected a small number of runs in which the initial BH-binary was randomly selected to contain a NS.  The second metric that we use is the {\edit number of BH+NS binaries a cluster is expected to contain at any given time}  $p_{\rm BH+NS} = N_{\rm BH} \tau_{\rm BH+NS}$.  In simulations with multiple BHs, $\tau_{\rm{BH+NS}}$ corresponds to the probability that any one of these BHs has a NS companion at a given time, and must therefore be scaled by $N_{\rm BH}$ to find {\edit the number of BH+NS binaries expected to exist in the cluster.}  The values of these metrics for each of our simulations are listed in \cref{tab:runs}.        

As was the case with the orbital parameter distributions, the structure of the background GC had a large impact on the values of \tbhns.  In cluster A ($n_{\rm c}=10^{4}~{\rm pc^{-3}}$), the BH+NS binaries were long-lived, with a mean lifetime of $7\times10^{9}$ yr.  However, because of the low encounter rate, these binaries were produced so rarely that \tbhns $\la 2 \times 10^{-3}$ in every simulation in this cluster.  Simulations in cluster D ($n_{\rm c}=10^{6}~{\rm pc^{-3}}$) had the opposite problem.  BH+NS binaries were formed efficiently in this cluster, but they were also rapidly driven to coalescence because of the high encounter rate.  The intermediate encounter rates in clusters B ($n_{\rm c}=10^{5}~{\rm pc^{-3}}$) and C ($n_{\rm c}= 5 \times 10^{5}~{\rm pc^{-3}}$) struck a balance between the production and destruction of BH+NS binaries. All other parameters held equal, simulations in clusters B or C had the largest values of \tbhns.  The BHs in these clusters typically spent a few percent of their lifetimes with a NS companion.

The amount of time that the BHs in our simulations had NS companions decreased with increasing $M_{\rm BH}$.  Both of the evolutionary processes that drive the BH+NS binaries to merge speed up when the mass of the BH is increased.  A more massive BH leads to a larger gravitational focusing cross-section, which increases the encounter rate. Furthermore, orbital decay is more rapid through gravitational wave emission when $M_{\rm BH}$ increases.  

{\edit Our models captured how the size and composition of a cluster's binary population will influence the production of BH+NS binaries.  The size of the background binary population was changed by varying the binary fraction.  Three different binary populations were considered, an observationally motivated population (OBS), a theoretically motivated population (FIR), and a population optimized to interact with the BHs (OPT; see \cref{sec:method} and CSC13 for a more complete description of these binary populations).}   The assumed background binary population did not influence \tbhns~in clusters A and B.  In these clusters mergers and disruptive encounters rarely occurred.  Accordingly, there were few single BHs in these clusters that needed to interact with the background binary population to acquire a new companion.  In clusters C and D, on the other hand, the nature of the background binary population had a substantial influence on \tbhns.  In both clusters it appears that there is a threshold around $f_{\rm b} \sim 0.2$.  Simulations with $f_{\rm b} < 0.2$ all exhibit similar values of \tbhns.  {\redit At larger $f_{\rm b}$, \tbhns~increased because single BHs were able to quickly exchange back into a binary}.  The composition of the binary population did not seem to impact the results much.  As we explained in CSC13, the main advantage of the FIR population over the OBS population is that it allows for larger binary fractions within observational constraints.  At the same value of $f_{\rm b}$, simulations using the FIR population and the OBS population produced BH+NS binaries with equal efficiency.   Use of the OPT population did, however, lead to larger \tbhns.~in clusters B, C, and D.                      

The simulations presented here also explored how changing the size of the NS and BH populations impacted the formation of BH+NS binaries.  The value of \tbhns~responded linearly to changes in the NS retention fraction ($f_{\rm ret}$).  In every simulation with $N_{\rm BH} > 1$, the increased size of the BH population lead to a reduction in the value of \tbhns.  When there was more than one BH in the simulation, it was common for the BH-binaries that we were evolving to interact with the other BHs in the cluster.  Nearly half of all three-body encounters between a BH+star binary and a second BH will result in the formation of a BH+BH binary \citep{Sigurdsson:1993a}.  {\edit The number of BH+BH binaries formed in each simulation is listed in \cref{tab:runs}.  In a future paper we will compute both the BH+BH and the BH+NS merger rates using our simulations with more than one BH.}   Once a BH+BH binary formed, it was nearly impossible for a NS to exchange into the system.  These BHs were essentially locked up for the rest of the run, leading to the reduction in \tbhns.  In addition to preventing the formation of BH+NS binaries, the presence of several BHs can also result in the destruction of BH+NS binaries.  In our simulations with multiple BHs, 10-50\% of BH+NS binaries were destroyed when another BH exchanged into the binary.  

For relatively small BH populations the value of \pbhns~increased compared to simulations with a single BH.  In these simulations the small decline in \tbhns~was outpaced by the increase in $N_{\rm BH}$.  However, the growth in \pbhns~quickly flattened out and/or turned over as $N_{\rm BH}$ was increased.  Comparing simulations 14, 15, and 16 shows that \pbhns~increased as we added more BHs to the cluster.  Extrapolating the trend seen in \tbhns~to larger values of $N_{\rm BH}$, we found that increasing $N_{\rm BH}$ to 100 would only boost {\edit the expected number of BH+NS binaries in this cluster} from 0.051 to 0.06.  Of course, one should use caution in drawing conclusions from such an extrapolation, but it seems unlikely that the presence of a substantial BH population would increase the size of a cluster's BH+NS binary population. In fact, such a large number of BHs could reduce the size of the BH+NS binary population.  This behavior is seen in some of our simulations.  In simulations 12 \& 13 and 31 \& 32 an increase in $N_{\rm BH}$ resulted in a reduction of \pbhns.  When there were many BHs in a cluster, the production of BH+BH binaries was favored over the production of BH+NS binaries.  

Combining all of the effects described above, we conclude that the probability of finding a BH+MSP binary is highest in massive GCs with $n_{\rm c}\sim{\rm few}~ \times~10^{5}\;{\rm pc^{-3}}$, $f_{\rm b} \ga 0.2$, and BH populations that comprise a few dozen $\sim 10\msun$ BHs.  We can estimate the number of BH+MSP binaries in the Milky Way GC system as:
\begin{equation}
N_{\rm BH+MSP} = N_{\rm GC}\;\;f_{\rm GC}\;f_{\rm MSP}\;f_{\rm BH}\;p_{\rm BH+NS},
\label{eqn:nbhmsp}
\end{equation}               
where, following \citet{Narayan:1991}, we have implicitly assumed that the lifetime of a MSP is $> 10^{10}$ yr.  Here $N_{\rm GC} = 150$ is the number of GCs in the Milky Way. The fraction of GCs with structural properties similar to those used in our simulations is denoted $f_{\rm GC}$.  Approximately $15-20\%$ of the Milky Way GCs have $n_{\rm c} \sim 10^{5}~{\rm pc^{-3}}$, $M_{\rm GC} = {\rm several} \times 10^{5}\msun$, and $1.7 < c_{\rm GC} < 2.0$ \citep{Harris:1996,Gnedin:1997}.  We gauged the fraction of NSs that have been recycled into MSPs, $f_{\rm MSP}$, using observational constraints on the total number of MSPs in 47 Tuc and Terzan 5.  \citet{Abdo:2010} used the integrated gamma-ray flux emitted by these clusters to estimate the number of MSPs, finding that 47 Tuc harbors $33^{+15}_{-15}$ MSPs and Terzan 5 contains $180^{+100}_{-100}$.   Using radio measurements, the total numbers of MSPs in 47 Tuc and Terzan 5 have been estimated to be $163^{+108}_{-70}$ and $294^{+224}_{-130}$, respectively  \citep{Chennamangalam:2013}.  Assuming that each of these clusters retains a total of 500-1000 NSs, we estimate that $f_{\rm MSP}$ is between 5\% and 30\% \citep{Pfahl:2002,Ivanova:2008}.  The fraction of GCs that retain at least one BH, $f_{\rm BH}$, is poorly constrained by observations. However, we can place an upper limit on the number of BH+MSP binaries produced in binary-single encounters by assuming that every massive GC retains a BH population. If we further assume the maximum reasonable value for every factor in \cref{eqn:nbhmsp}, we find that the upper limit for the number of BH+MSP binaries produced through this channel in the Milky Way GC system is $\sim10$.  Here we have used the value of \pbhns~$=1.1$ computed in simulation 28.  The number of detectable BH+MSP binaries is a factor of 2--3 smaller due to beaming effects.  The upper limit on the number of dynamically formed BH+MSP binaries presented here is similar to the number of BH+MSP binaries expected to form through the evolution of isolated binaries \citep{Sipior:2004,Pfahl:2005}.  

{\edit We construct our best estimate for} the total number of BH+MSP binaries by generating several Monte Carlo realizations of the Milky Way GC population and counting the number of BH+MSP binaries in each realization.  In these models we used $N_{\rm GC} = 150$.  The mass of each cluster was drawn from the GC mass function presented in \citet{McLaughlin:1996}.  We then assigned each cluster a \pbhns~by randomly selecting one of our simulations that was done in a GC of similar mass.  Next, we chose $f_{\rm MSP}$ for each cluster from a normal distribution with a mean of 0.13 and a standard deviation of 0.07.  {\edit We chose to cast our estimates in terms of the fraction of GCs that retain at least one BH ($f_{\rm BH}$) because this quantity is so poorly constrained.}  Based on $10^{4}$ realizations, we found $N_{\rm BH+MSP} = (0.6\pm0.2) f_{\rm BH}$.  In computing this number we assumed that the size of each GC's BH population was random.  It is possible that the sizes of the BH populations are correlated, i.e., the GCs that retain BHs either all retain several or all retain $\sim 1$.  If we recompute our Monte Carlo realizations and require that $N_{\rm BH} >1$, then we find $N_{\rm BH+MSP} = (1.3\pm0.3) f_{\rm BH}$.  If we only consider simulations that used $N_{\rm BH} =1$, the size of the BH+MSP binary population is reduced to  $N_{\rm BH+MSP} = (0.2\pm0.1) f_{\rm BH}$.  We have not accounted for the fact that many components of these estimates, (e.g.,  $N_{\rm BH}$, $f_{\rm BH}$, and $f_{\rm MSP}$) are likely to be functions of the GCs' structural parameters.  However, it is unlikely that including these dependencies will significantly alter our conclusion that $N_{\rm BH+MSP} \la 1$.      {\edit This estimate suggests that dynamically formed BH+MSP binaries in GCs may be even rarer than those that are likely to be produced through standard binary evolution in the disk of the Galaxy.}                   

\section{Discussion and Conclusions}
\label{sec:discussion}
We have presented a study of the dynamically formed BH+MSP binaries in GCs.  We found that in the highest density clusters ($n_{\rm c} \ga 5 \times 10^{5}~{\rm pc^{-3}}$), the semimajor axis distribution of the BH+MSP binaries is nearly independent of all of the parameters that we varied in our study.  This property of the BH+NS binary populations is {\edit beneficial for observers} who hope to identify such systems.   Regardless of the nature of many uncertain characteristics, including the GC BH and binary populations, the vast majority of BH+MSP binaries produced in dense GCs will have $2 < P_{\rm B} < 10$ days.  In lower density clusters, $M_{\rm BH}$ does influence the expected orbital periods of the BH+MSP binaries.  In clusters with $n_{\rm c} \sim 10^{5}~{\rm pc^{-3}}$ BH+MSP binaries with {\edit massive BHs} ($M_{\rm BH} = 35\msun$) will typically have orbital periods around 20 days.  For BH+MSP binaries with $7\msun$ BHs, the expected orbital periods are much longer, with typical periods in the 150 to 250 day range. 

{\edit Importantly, we have also found} that dynamically formed BH+MSP binaries are quite rare. We estimated that the maximum number of detectible BH+MSP binaries produced through this channel in the Milky Way GC {\edit system is  approximately 3--5}.  Comparing the size of the BH+MSP binary population predicted by our models to population synthesis models of such binaries in the field, we find that the dynamical encounters result in a factor of $\sim$100 enhancement in BH+MSP binary production per unit mass in GCs.  The birthrates of other exotic objects (e.g., LMXBs, MSPs) in GCs receive a similar boost over the field due to the additional, dynamical formation channels available to the members of a dense stellar system.        

The small size of the population is not a consequence of our assumption that most stellar mass black holes are ejected from the cluster early in its evolution.  The presence of {\edit many BHs will} also reduce the probability that a cluster harbors a BH+MSP binary.  BH+MSP binary formation can be stifled by as few as 19 BHs.  If there are several BHs in the cluster the BHs will preferentially interact with each other and not the NSs.  Furthermore, any BH+NS binaries that are formed may be destroyed when another BH exchanges into the binary.  This behavior has also been seen in models that considered the evolution of the BH population as a whole. \citet{Sadowski:2008} and \citet{Downing:2010} found that very few BH+NS binaries were produced in their simulations, which included several hundred to over one thousand BHs.  We expect dynamically formed BH+MSP binaries to be rare regardless of the size of the retained BH population.

Another factor that played a surprisingly small role in limiting the size of the BH+MSP binary population were the large post merger recoil that BH+NS binaries are expected to receive.  Several recent studies have used numerical relativity to simulate BH+NS binary mergers, and these models show that the remnant BH will receive a kick of more than $50$ km s$^{-1}$ when $3\la M_{\rm BH}/M_{\rm NS} \la 10$ \citep[e.g.][]{Shibata:2009,Etienne:2009,Foucart:2011}.  These kicks exceed the escape velocities of all but the most massive GCs, so BHs of $\sim 7 \msun$ will be ejected from the cluster once they merge with a NS.  These post-merger ejections reduce \tbhns because they act to remove single BHs from the cluster.  However, there are three ways to avoid the large recoils.  First, at smaller mass ratios, the NS is tidally disrupted before the merger, which halts the anisotropic emission of gravitational radiation and suppresses the kick.  Unfortunately, it is unlikely that a BH+NS binary with such a small mass ratio exists in nature. Second, the linear momentum flux responsible for the kick declines for larger values of  $M_{\rm BH}/M_{\rm NS}$, again reducing the magnitude of the kick \citep[see, e.g.][]{Fitchett:1983}.  Finally, a large BH spin could also decrease the magnitude of the post merger kick.  \citet{Foucart:2013} showed that for BHs with dimensionless spin parameters of 0.9, the recoil would be smaller than typical GC escape velocities for BH+NS binaries with mass ratios as small as 7.  We tested how these latter two, plausible scenarios would affect our results.

Many of the models discussed above already illustrate the case in which the post merger kick is suppressed because $M_{\rm BH} \gg M_{\rm NS}$.  In the simulations that used $35 \msun$ BHs, the kick is small enough that most of the BHs are retained by the cluster after merging with a NS.  However, as we previously discussed in \cref{sec:detection}, the BH+NS binaries produced in these models do not live as long as the binaries produced in simulations with lower mass BHs.  The presence of a more massive BH accelerates both of the hardening mechanisms that drive the BH+NS binaries to merge.  We also tested a more extreme mass ratio by running a simulation with a $100 \msun$ BH.  We found \tbhns = 0.036 in this model, which is similar to the value of \tbhns for a simulation in the same background cluster with a 35 \msun BH.  Furthermore, the IMBH+MSP formation rate implied by this value \tbhns is consistent with previous work on the formation of such binaries by \citet{Devecchi:2007}.   Increasing the mass of the BH to suppress the kick does not significantly increase \tbhns, and accordingly the size of the BH+MSP binary population.             

To test the high spin scenario, we reran simulations using one 7 \msun BH in the high density clusters ($n_{\rm c} \ge 5 \times 10^{5}$ pc$^{-3}$) with the post merger kicks switched off.  In both cases \tbhns increased by nearly a factor of two, to 0.13 and 0.08 in clusters C and D, respectively.  Despite the significant increase in \tbhns, we still find $N_{\rm BH+MSP} \la 1$ in the Milky Way GC system.  It should also be noted that these new simulations actually overestimate the number of rapidly spinning BHs retained by the GCs. The kick is only reduced significantly if the misalignment between the BH's spin and the angular momentum of the BH+NS binary is $\la 60\degr$.  For dynamically formed binaries, it is likely that the orbital angular momentum and the spins of their components will have random orientations.  Furthermore, we also counted the number of times the BHs in these simulations merged with non-NSs to estimate the degree to which the BHs would be spun up by thin disk accretion.  {\edit We will discuss these results more broadly in the context of X-ray binary production in a future publication.  For now we will only examine whether the BHs are able to accrete a substantial amount of angular momentum.}  In both of the background clusters considered, the BHs merged with an average of 0.97 non-NSs during their lifetimes.  It was only in rare cases the BHs merged with 7-9 stars. Thus, it is unlikely that the BH's spin will increase substantially during its evolution in the cluster.  The BHs must be born with large spins  for this post merger recoil mechanism to be effective.  Even if this is the case, we expect that BH+MSP binaries will be extremely rare in GCs.                      

Some limitations of our method will impact the results of our simulations.  Because our simulations do not include binary--binary encounters, they do not capture several processes that affect the formation of BH+MSP binaries.  As discussed above,  binary--binary interactions open up additional BH+NS binary formation channels.  Furthermore, in clusters with multiple BHs, collisions between pairs of BH+BH binaries could eject or disrupt many BH+BH binaries \citep[e.g.][]{OLeary:2006,Banerjee:2010,Downing:2010,Tanikawa:2013}. Reducing the number of BHs in the cluster and freeing BHs from otherwise impenetrable BH+BH binaries would increase the likelihood that BH+MSP binary is produced.  However, binary--binary interactions will also disrupt and eject BH+MSP binaries.  Models that include binary--binary interactions are needed to see which processes dominate.  

{\redit We also neglected long-range interactions between BH-binaries and background stars.  These interactions do no perturb the binary's orbital parameters as strongly as close encounters, but they do occur more frequently.  Of particular concern is whether the change in eccentricity resulting from these encounters will accelerate the rate of orbital contraction through the emission of gravitational waves (see \cref{eqn:dadtgw}). For eccentric binaries, the change in eccentricity induced by an encounter declines as $r_p^{-3/2}$, where $r_p$ is separation between the binary and single star at pericenter \citep[][]{Heggie:1975}.  The change in eccentricity declines even faster with increasing $r_p$ for circular binaries  \citep{Hut:1984, Rappaport:1989,Phinney:1992,Rasio:1995}. These encounters drive a random walk in eccentricity because they are just as likely to increase a binary's eccentricity as they are to decrease it.  We used the cross-sections for eccentricity change derived by \citet{Heggie:1996} to estimate the root mean square rate of change in eccentricity induced by the distant encounters that our models did not include.  Given these rates, we found that the eccentricity of the binaries in our simulations would  change by $< 0.05$ over their entire lifetimes.  In simulation 10 ($n_{\rm c} = 10^5$ pc$^{-3}$, one 7\msun BH), for example,  the median eccentricity change amongst the BH+NS binaries was 0.008.  Allowing for this modest change in eccentricity changed the median gravitational wave merger time for the BH+NS binaries by $\pm 7\%$.  The median merger time for BH+WD binaries changed by $\pm 9\%$.  Distant encounters will only have a small effect on the BH+MSP binary population.}

{\redit Finally, as a consequence of the assumption that the background cluster was static, our simulations were unable to capture some  dynamical processes.  In the static cluster models,} we forced the BHs to remain in equilibrium with the rest of the cluster.  If we had allowed for the dynamical evolution of the BHs, they might have decoupled from the cluster and produced a dense, inner subcluster.  This would have further reduced the number of encounters between BHs and NSs.  {\edit Alternatively, heating of the cluster by the BHs could result in expansion of the core, which would result in longer lifetimes for any BH+MSP binaries that managed to form \citep{Mackey:2008,Heggie:2014}.  {\redit Primordial binaries are an additional source of heating that we were unable to included in our models.  We showed that efficient BH+MSP binary formation required substantial binary fractions, but we did not account for the impact heating by these large binary populations could have on the structure of the cluster.}    Clearly, more detailed models are needed to determine how these additional processes impact the BH+MSP populations in GCs.}                                
    
Although the number of BH+MSP binaries in the Milky Way GC system is expected to be small, searching for these binaries is still warranted.  We know that these binaries are in GCs, and our models make specific predictions about the types of GCs that are likely to harbor BH+MSP binaries.  Given the rarity of BH+MSP binaries predicted by our models, the discovery of a BH+MSP pulsar binary in the Milky Way GC system would imply that the fraction of clusters that retain at least one stellar mass BH is large.  With the potential scientific payoff, continued deep radio observations of the cores of the $\sim 20$ Milky Way GCs with appropriate structural properties may be justified.  The clusters most likely to host a BH+MSP binary include 47 Tuc, Terzan~5, NGC 1851, NGC 6266, {\edit NGC 6388} and NGC 6441.  {\edit Intriguingly, previous theoretical and observational studies have suggested that NGC 6388 and NGC 6441 may harbor BHs \citep{Lanzoni:2007,Moody:2009}.} Finally, even though there might not be any BH+MSP binaries in the Milky Way GC system, such binaries could be detected in extra-galactic GCs with the Square Kilometer Array (SKA).  SKA should be able to detect most pulsars within 10 Mpc \citep{Cordes:2007}, and our models predict there could be $\sim 100$ dynamically formed BH+MSPs binaries within this volume.    

\bigskip
{\noindent {\bf ACKNOWLEDGMENTS}}

\noindent The authors thank E.S. Phinney, Francois Foucart, and  Lawrence E. Kidder for valuable discussion. We also thank the referee for an insightful review that helped us improve the paper. SS and DFC thank the Kavli Institute for Theoretical Physics for their hospitality.  SS thanks the Aspen Center for Physics.  DC acknowledges support from NSF grant AST-1205732.

\bibliography{all_refs}

\end{document}